\documentclass{iopconfser}

\usepackage{amsmath}
\usepackage{graphicx}
\usepackage{amssymb}
\usepackage{hyperref}
\usepackage{float}
\usepackage{xcolor}

\begin{document}

\title{Algebraic classification of 2+1 geometries}

\author{M Papaj\v{c}\'ik and J Podolsk\'y}

\affil{Institute of Theoretical Physics, Charles University, Prague, Czechia}

\email{matus.papajcik@matfyz.cuni.cz}

\begin{abstract}
We present a new effective method of algebraic classification of 2+1 geometries. Our approach simply classifies spacetimes using five real scalars, defined as specific projections of the Cotton tensor onto a suitable null basis. The algebraic type of a spacetime is determined by gradual vanishing of these scalars. We derive the Bel--Debever criteria, together with the multiplicity of the Cotton aligned null directions (CANDs). Additionally, we provide a frame-independent algorithm for classification based on the polynomial curvature invariants and show the equivalence to previous methods of classification.
\end{abstract}

\section{Introduction}

Gravitation in three dimensions (3D) provides a simplified setting in which some conceptual ideas of physics can be explored. For example, quantum gravity offers a rich area of research in such a context. This is largely due to Witten, who showed that 3D Einstein gravity can be formulated as a Chern--Simons theory \cite{Witten:1988} and is exactly quantizable and renormalizable. Many subsequent works developed additional approaches to quantum gravity and a review of the subject is available in the monograph \cite{Carlip:1998}. Moreover, in general relativity a four-dimensional (4D) spacetime admitting a translational symmetry is equivalent to a system of Einstein's equations with a massless scalar field in 3D. This well-known property and the related symmetries were studied in detail by Geroch \cite{Geroch:1971}. Consequently, there has been a growing interest in exact solutions of 3D general relativity and its modifications. Nowadays, a large number of exact solutions are known, see the monograph \cite{Garcia-Diaz:2017} for their overview. Therefore, a classification of these solutions requires a robust method, such as the Newman--Penrose formalism \cite{Penrose:1960} used in 4D. In this contribution, we summarize our recent results \cite{PP:2023,PP:2024}, where we introduce a new approach to algebraic classification of 3D spacetimes.

\section{New approach to algebraic classification in 3D}

Local curvature in 3D is fully determined by the Ricci tensor and the Ricci scalar. The Weyl tensor identically vanishes and therefore classification methods known from $D \geq 4$ must be significantly altered. This obstacle can be resolved by employing a different conformally invariant tensor in 3D, derived already by Cotton in 1899 \cite{Cotton:1899}. It serves a role analogous to the Weyl tensor, for example it also vanishes for conformally flat spacetimes, and in 3D it is defined as 
\begin{equation} \label{Cotton_tensor}
C_{abc} \equiv 2 \Big( \nabla _{[a}R_{b]c}-\tfrac{1}{4}\nabla _{[a}R \, g_{b]c}\Big) \, ,
\end{equation} 
where $R_{ab}$ is the Ricci tensor and $R$ is the Ricci scalar. At any point, in a given spacetime, we can construct a real null frame ${\{ \mathbf{k},\, \mathbf{l},\, \mathbf{m} \} }$, consisting of two null vectors ${\mathbf{k} \cdot \mathbf{k} = 0 = \mathbf{l} \cdot \mathbf{l}}$ and a vector in the remaining transverse spatial dimension ${\mathbf{k} \cdot \mathbf{m} = 0 = \mathbf{l} \cdot \mathbf{m}}$, satisfying the normalization conditions
\begin{equation} \label{Null_frame_normalization}
\mathbf{k} \cdot \mathbf{l} = -1 \qquad \mathbf{m} \cdot \mathbf{m} = 1 \, .
\end{equation}
Projecting the Cotton tensor \eqref{Cotton_tensor} onto this null frame, we obtain five real scalars that encode its five independent components. Specifically, we define the following scalars of the Cotton tensor, see Eq. (2.3) in \cite{PP:2023} or Eq. (9) in \cite{PP:2024}
\begin{align}
\Psi _0 &\equiv C_{abc} \, k^a \, m^b \, k^c  \, ,&
\Psi _1 &\equiv C_{abc} \, k^a \, l^b \, k^c  \, ,&
\Psi _2 &\equiv C_{abc} \, k^a \, m^b \, l^c \, ,&\label{Cotton_scalars}\\
\Psi _3 &\equiv C_{abc} \, l^a \, k^b \, l^c \, ,&
\Psi _4 &\equiv C_{abc} \, l^a \, m^b \, l^c \, . \nonumber
\end{align}
Similarly to the Newman--Penrose classification in 4D, the spacetimes are classified into five algebraic types according to the vanishing of the Cotton scalars \eqref{Cotton_scalars}. The specific conditions satisfied by each algebraic type are given in Table~\ref{Table_classification}.
\begin{table}[t]
\begin{center}
\begin{tabular}{ c c l l} 
\hline
\\[-8pt]
algebraic type && \qquad\quad the conditions\\[2pt]
\hline
\\[-8pt]
I   && ${\Psi _0=0}$\,,                              & $\Psi _1\ne0$ \\[2pt]
II  && ${\Psi _0=\Psi _1=0}$\,,                      & $\Psi _2\ne0$ \\[2pt]
III && ${\Psi _0=\Psi _1=\Psi _2=0}$\,,              & $\Psi _3\ne0$ \\[2pt]
N   && ${\Psi _0=\Psi _1=\Psi _2=\Psi _3=0}$\,,      & $\Psi _4\ne0$ \\[2pt]
D   && ${\Psi _0=\Psi _1 = 0 = \Psi _3=\Psi _4}$\,,  & $\Psi _2\ne0$ \\[2pt]
\hline
\end{tabular}
\caption{Algebraic classification of 2+1 geometries given by the Cotton scalars \eqref{Cotton_scalars}.}
\label{Table_classification}
\end{center}
\end{table}

\section{Cotton aligned null directions (CANDs)}

\begin{table}[b]
\begin{center}
\resizebox{11.5cm}{!}{
\begin{tabular}{cccll}
\hline
\\[-8pt]
  algebraic type & CANDs & {multiplicity\quad} & & \hspace{-36mm} canonical Cotton scalars \\[2pt]
\hline
\\[-8pt]
   I
   & \hbox{
   \rotatebox[origin=c]{-30}{$\leftarrow$}\hspace{-3mm}
   \raisebox{1.5mm}{\rotatebox[origin=c]{-60}{$\leftarrow$}}\hspace{-1mm}
   \raisebox{1.5mm}{\rotatebox[origin=c]{60}{$\rightarrow$}}\hspace{-3mm}
   \rotatebox[origin=c]{30}{$\rightarrow$}}
   & \hspace{-1mm}${1+1+1+1\quad}$
   & ${\Psi_0=0}$\,,
   & ${\Psi_1\ne0}$     \\[2pt]
   II
   & \hbox{
   \rotatebox[origin=c]{-30}{$\leftarrow$}\hspace{-3mm}
   \raisebox{1.5mm}{\rotatebox[origin=c]{-60}{$\leftarrow$}}\hspace{-1mm}
   \raisebox{0.6mm}{\rotatebox[origin=c]{45}{$\Rightarrow$}}}
   & ${1+1+2\quad}$
   & ${\Psi_0=\Psi_1=0}$\,,
   & ${\Psi_2\ne0}$    \\[6pt]
   D
   & \hbox{
   \rotatebox[origin=c]{-45}{$\Leftarrow$}\,\rotatebox[origin=c]{45}{$\Rightarrow$}}
   & ${2+2\quad}$
   & ${\Psi_0=\Psi_1=0=\Psi_3=\Psi_4}$\,,
   & ${\Psi_2\ne0}$    \\[4pt]
   III
   & \hbox{
   \rotatebox[origin=c]{-30}{$\leftarrow$}\hspace{-1mm}
   \raisebox{0.6mm}{\rotatebox[origin=c]{45}{$\Rrightarrow$}}}
   & ${1+3\quad}$
   & ${\Psi_0=\Psi_1=\Psi_2=0}$\,,
   & ${\Psi_3\ne0}$     \\[6pt]
   N
   & {\Large
   \hbox{
   \rotatebox[origin=c]{45}{$\Rightarrow$}\hspace{-6.1mm}
   \raisebox{-0.30mm}{\rotatebox[origin=c]{45}{$\Rightarrow$}}}
   }
   & $4\quad$
   & ${\Psi_0=\Psi_1=\Psi_2=\Psi_3=0}$\,,
   & ${\Psi_4\ne0}$     \\[2pt]
\hline
\end{tabular}
}
\caption{Relationship between algebraic types, CANDs and corresponding multiplicity of roots of the equation \eqref{CAND_quartic}.}
\label{Table_CAND}
\end{center}
\end{table}

To determine the algebraic type of a spacetime, we reformulate the classification based on the Cotton scalars into a set of useful relations. In fact, the specific conditions in Table~\ref{Table_classification} are equivalent to the existence of a special null vector field $\mathbf{k}$. These relations are known as the Bel--Debever criteria and in 3D they have the form
\begin{align}
k_{[d}\,C_{a]bc} \, k^b \, k^c &=0 \quad \Leftrightarrow \quad \Psi _0=0 \, , &
C_{abc} \, k^b \, k^c &=0 \quad \Leftrightarrow \quad \Psi _0=\Psi _1=0 \, ,  \\
k_{[d}\,C_{a]bc} \, k^b &=0 \quad \Leftrightarrow \quad \Psi _0=\Psi _1=\Psi _2=0 \, ,&
C_{abc} \, k^b &=0 \quad \Leftrightarrow \quad \Psi _0=\Psi _1=\Psi _2=\Psi _3=0 \, .\nonumber
\end{align}
Finding this null direction can be achieved by using the freedom in the choice of the null basis satisfying \eqref{Null_frame_normalization}. Such bases can only differ by local Lorentz transformations. In view of the classification method only the following subgroups are important
\begin{align}
\text{boost: }& & \mathbf{k}^{\prime} &= B\, \mathbf{k}\, , & \mathbf{l}^{\prime} &= B^{-1}\, \mathbf{l} \, , & \mathbf{m}^{\prime} &= \mathbf{m} \, ,\label{Boost}\\
\text{null rotation: }& & \mathbf{k}^{\prime} &= \mathbf{k} + \sqrt{2}K \, \mathbf{m} + K^2 \mathbf{l}\, , & \mathbf{l}^{\prime} &= \mathbf{l} \, , & \mathbf{m}^{\prime} &= \mathbf{m}+\sqrt{2}K\, \mathbf{l} \, .\label{Null_rotation}
\end{align}
Applying the boost \eqref{Boost}, the Cotton scalars \eqref{Cotton_scalars} transform as ${\Psi'_{\rm A} = B^{2-{\rm A}}\,\Psi_{\rm A}}$. This means that the Cotton scalars are ordered such that the scalars gradually vanish, starting from the highest boost weight (power of the parameter $B$). This is in full agreement with the classification of any tensor in any dimension developed in \cite{MCPP:2005}. The null rotation with fixed $\mathbf{l}'=\mathbf{l}$ \eqref{Null_rotation} allows us to find the Cotton aligned null direction $\mathbf{k}'$ for which $\Psi _0'=0$, see Eq. (45) in \cite{PP:2024}. This leads to a polynomial equation of the fourth order
\begin{equation} \label{CAND_quartic}
\Psi_4\,K^4 - 2\sqrt{2}\,\Psi_3\,K^3 - 6\,\Psi_2\,K^2 + 2\sqrt{2}\,\Psi_1\,K  - \Psi_0 =0 \, .
\end{equation}
The number of independent CANDs in a spacetime determines the algebraic type. This is directly related to the number of roots, and their multiplicity, of the equation \eqref{CAND_quartic}, see Table~\ref{Table_CAND} for details. 

\section{Polynomial invariants}

\begin{figure}[H]
\begin{center}
\includegraphics[scale=0.65]{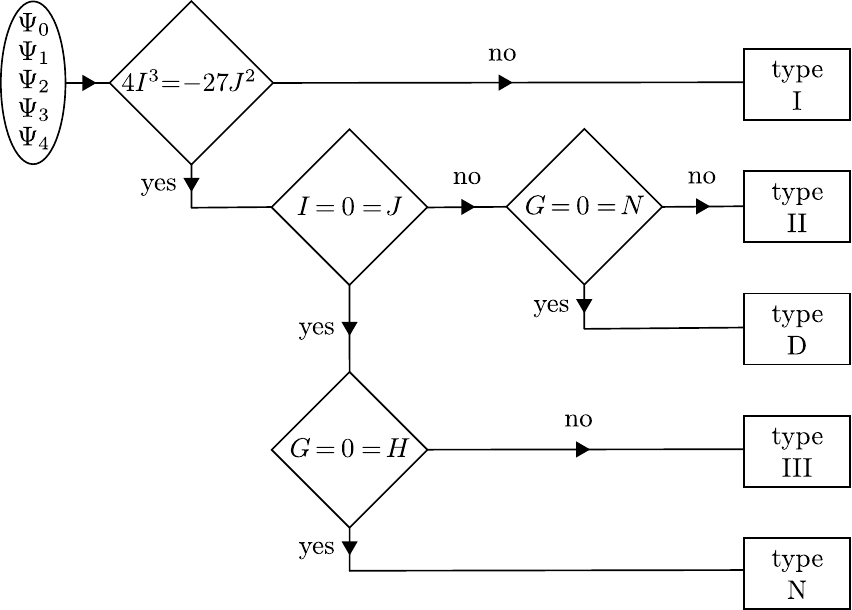}
\end{center}
\caption{Flow diagram for determining the algebraic type of 2+1 geometry using polynomial invariants.}
\label{Flow_diagram}
\end{figure}

A frame-independent classification method can be constructed from invariants of the Cotton tensor \eqref{Cotton_tensor} and the Cotton--York tensor \cite{York:1971}, which is defined as its Hodge map
\begin{equation} \label{Cotton-York_tensor}
Y_{ab} \equiv \tfrac{1}{2} \, g_{ak}\, \epsilon^{kmn}  \, C_{mnb} \, ,
\end{equation}
where $\epsilon^{kmn}$ is the Levi--Civita tensor. There are two main invariants defined as
\begin{align}
I &\equiv \tfrac{1}{4}C_{abc}\,C^{abc} =  \Psi _0 \Psi _4 -2\,\Psi _1 \Psi _3 -3\,\Psi _2^2\, ,\\
J &\equiv \tfrac{1}{6}C_{abc} \,C^{abd}\, {Y^c}_{d} = \Psi _0\Psi _3^2 -\Psi _1^2\Psi _4 +2\,\Psi _0\Psi _2\Psi _4 +2\,\Psi _1\Psi _2\Psi _3 +2\,\Psi _2^3 \, .
\end{align}
Additionally, we introduce the following polynomial covariants for a complete analysis
\begin{align}
G \equiv \Psi _1\Psi _4^2 -3\,\Psi _2\Psi _3\Psi _4 -\Psi _3^3 \, ,&&
H \equiv 2\,\Psi _2\Psi _4 + \Psi _3^2 \, ,&&
N \equiv 3\, H^2 + \Psi _4^2\, I  \, .
\end{align}
These expressions occur naturally in the analysis of the roots of the quartic equation \eqref{CAND_quartic}. Algebraic type of a spacetime is then determined by simple conditions. The corresponding algorithm for determining the type from Cotton scalars \eqref{Cotton_scalars} evaluated in any basis is given in Figure~\ref{Flow_diagram}. 
It should be noted that in this formalism we allow complex solutions of equation \eqref{CAND_quartic} which leads to complex CANDs. This can only happen for type~I when $4I^3>-27 J^2$ where there are two complex conjugate CANDs, or if this relation is not satisfied and $H<0$ or $N<0$, in which case there are four complex CANDs. For type~II there are complex conjugate CANDs if $N<0$ and for geometry of type~D there are two conjugate complex CANDs of multiplicity two if $H<0$.

Using the method of invariants, we can establish the relation to Petrov-like classification of the Cotton--York tensor \eqref{Cotton-York_tensor} originally proposed in \cite{BBL:1986} and later refined in \cite{GHHM:2004}. This approach to algebraic classification is based on the eigenvalue problem of the Cotton--York tensor
\begin{equation}
Y{_a}^{b}\, \nu _b = \lambda \, \nu _a \quad \Leftrightarrow \quad \text{det}(Y{_a}^{b}-\lambda \, \delta {_a}^b)=0 \, .
\end{equation}
The algebraic types are then determined by the specific Jordan decomposition of the matrix $Y{_a}^{b}$. Notably, the classification in terms of the Cotton scalars is fully equivalent to the Petrov-like approach in 3D.

\section*{Acknowledgments}

This work was supported by the Czech Science Foundation Grant No.~GA\v{C}R 23-05914S and by the Charles University Grant No.~GAUK 260325.

\end{document}